\documentclass[11pt,twoside]{article}
\usepackage{macro-fsut-eng}
\usepackage{psfig}
\usepackage{graphicx}

\usepackage{amsmath,amssymb}

\usepackage[T1]{fontenc} 

\usepackage{latexsym}
\usepackage{verbatim}

\begin{document}

\vskip 1.0cm
\markboth{M.~Sinha \& Holley-Bockelmann}{Flybys in the Universe}
\pagestyle{myheadings}

\vspace*{0.5cm}
\title{A First Look at Galaxy Flyby Interactions: Characterizing the Frequency of Flybys in a Cosmological Context}

\author{M.~Sinha$^1$ and  K.~Holley-Bockelmann$^1$}
\affil{$^1$Department of Physics \& Astronomy, Vanderbilt University, Nashville, TN 37235, USA}

\begin{abstract}
Hierarchical structure formation theory is based on the notion that mergers drive galaxy evolution, 
so a considerable framework of semi-analytic models and N-body simulations has been constructed to 
calculate how mergers transform a growing galaxy. However, galaxy mergers are only one type of major 
dynamical interaction between halos -- another class of encounter, a close flyby, has been  largely ignored. 
We analyze a 50 $h^{-1}$ Mpc, $1024^3$ collisionless cosmological simulation and find that the number of close flyby interactions is 
comparable to, or even surpasses, the number of mergers for halo masses $\ga 10^{11}\,{h^{-1} M_\odot}$ at $z \la 2$. Halo 
flybys occur so frequently to high mass halos that they are continually perturbed, unable to reach a dynamical equilibrium.
We also find tentative evidence that at high redshift, $z \ga 14$, flybys are as frequent as mergers. 
Our results suggest that close halo flybys can play an important role in the evolution of the earliest dark matter halos and their 
galaxies, and can still influence galaxy evolution at the present epoch.
\end{abstract}

\section{Introduction}
In a $\Lambda$CDM Universe, the smallest dark matter halos form first; bigger halos are then formed
via successive mergers with smaller halos. Thus, mergers are instrumental in the formation and evolution
of halos.  Mergers can dramatically change a galaxy -- from its morphology (e.g., Holmberg 1941, Toomre \& Toomre 1972, Barnes 2002), 
to its stellar population (e.g., Mihos \& Hernquist 1994), 
to the evolution of the central supermassive black hole~(e.g., Hopkins et al.~2006, Micic et al.~2011). 
Consequently, merger rates have been studied extensively, both  theoretically~(e.g., Lacey \& Cole 1993, Genel et al.~2009)
and observationally~(e.g., Schweizer 1986). Collisionless cosmological N-body
simulations can be used to measure halo merger rates, where a merger is defined to occur when a bound dark matter halo
falls into another bound dark matter halo. 
Galaxy merger rates can then be inferred from the subhalo mergers within a primary halo by assuming a $M_{\rm halo} - M_{\rm gal}$ 
relation~(e.g., Guo \& White 2008, Wetzel et al.~2009) or directly measured in hydrodynamic simulations~(e.g., Maller et al.~2006, Simha et al.~2009).
Observationally, galaxy merger rates are typically derived from close-pair counts -- i.e. galaxies with small projected separations 
and relative velocities -- and are globalized using an estimate of the lifetime or duration of the observed merger phase~(Lotz et al. 2010, Zavala et al. 2012).

Ultimately, galaxy mergers are successful in shaping galaxy properties
because they cause a large perturbation within the potential. 
However, one entire class of galaxy interactions
also capable of causing such perturbations -- galaxy flybys -- has
been largely ignored. Unlike galaxy mergers where two galaxies combine into one remnant, flybys 
occur when two independent galaxy halos interpenetrate 
but detach at a later time; this can generate a rapid and large perturbation in each galaxy. We developed and tested a method to 
identify mergers and flybys between dark matter halos in cosmological simulations and to construct a full `interaction network' 
that assesses the past interaction history of any given halo. In this work, we present a census of halo flybys and mergers in the Universe determined
from cosmological simulations. Please see Sinha \& Holley-Bockelmann (2012) for more details. 

\section{Methods}
We use a high-resolution, dark matter simulation with $1024^3$ particles in a box of length $50 h^{-1}$ Mpc
with {\small WMAP-5} cosmological parameters as a testbed to develop our technique.  We 
use a fixed, co-moving softening length of $2.5 h^{-1}$ kpc and evolve the particles 
with {\small GADGET-2}~(Springel et al.~2001a, Springel 2005). 
Since the fundamental mode goes non-linear at $z=0$, we will only present results up to $z=1$ where the $50 h^{-1}$ Mpc
box is still a representative cosmological volume. 
To begin identifying halos, we first use a Friends-of-Friends({\small FOF}) technique with a
canonical linking length $b=0.2$ ($\sim 10 h^{-1}$ kpc).
We require at least 20 particles  ($\sim 10^8 \,{h^{-1} M_\odot}$) to define a halo, but our halo
interaction network uses only those halos with greater than 100 particles. Subhalos are
identified using the {\small SUBFIND} algorithm~(Springel et al. 2001b). 

We classify a flyby using spatial {\em and} dynamical information. Conceptually, a grazing flyby occurs when a halo
undergoes the transition from a main halo $\boldsymbol\rightarrow$subhalo$\boldsymbol\rightarrow$ a main halo and is fundamentally different
from mergers. However, using only this main halo $\boldsymbol\rightarrow$subhalo$\boldsymbol\rightarrow$ a main halo transition would 
also capture spurious interactions where one halo just skirts the outer edge of another halo.  We eliminate these spurious 
interactions by only choosing the flybys that have a duration longer than half 
the crossing time\footnotemark. We find that with this definition the rate of flybys converges irrespective of the frequency of 
snapshot outputs. In a simulation, it is relatively straightforward to identify such chains since the current kinematics
and future behavior of a given interaction is fully determined. 
We note that because flybys imply a population of galaxies that were once well within
the virial radius of the main halo but are now outside it, they may be related to the so called `backsplash' galaxies~(Gill et al.~2005).

\footnotetext{Since the skirting encounters would necessarily be very short-lived. By imposing this minimum duration, we 
are capturing the target orbit, i.e., that of a subhalo going through a main halo.}

\section{Results}
\begin{figure} 
\centering
\includegraphics[width=0.55\textwidth,keepaspectratio=true,clip=true]{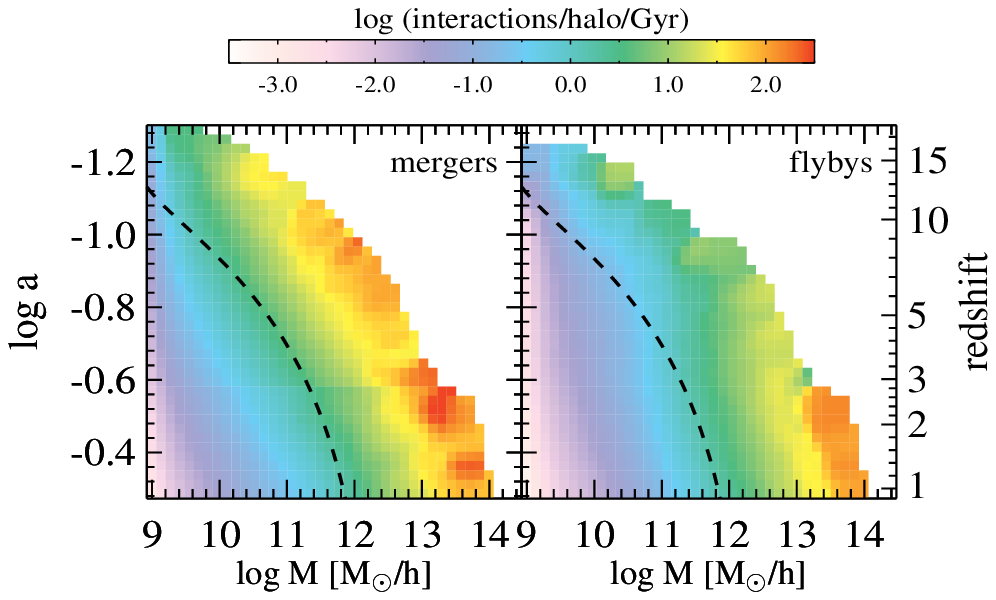}%
\includegraphics[width=0.45\textwidth,keepaspectratio=true,clip=true]{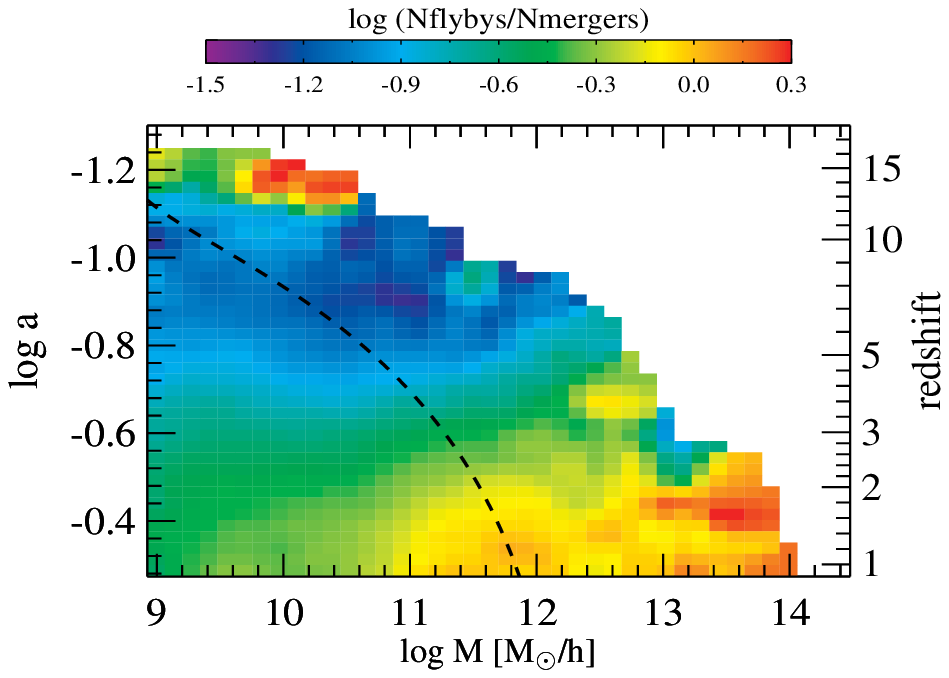}%
\caption{\small {\em (Left).} This figure shows the number of mergers (left) and flybys (right) per halo per Gyr as a function of primary halo mass and redshift. 
The dashed line shows the mass accretion history of a typical Milky-Way type obtained from our simulations. The number
of flybys increases with the primary halo mass for all redshifts, consistent with $\Lambda$CDM. For $z\la 3$, halos above $10^{13}\,{h^{-1} M_\odot}$ 
have flyby rates greater than 100 per Gyr. Such a high flyby rate (in addition to mergers) means these halos are unlikely to be in
equilibrium. {\em (Right).} Ratio of number of flybys to mergers as a function of primary halo mass and redshift. Mergers dominate flybys 
by an order of magnitude for $12\la z\la 4$. At lower redshifts, however, flybys start becoming
more prevalent and by $z\sim 2$, flybys are at comparable or even larger for all halos above $10^{11} \,{h^{-1} M_\odot}$.  }
\label{fig:rate_interactions_gyr_primary}
\end{figure}

In the left panel of Figure~\ref{fig:rate_interactions_gyr_primary}
we show the number of flybys and mergers on a per halo per Gyr basis, while the ratio of flybys to mergers is
seen in the right panel.  We see that flybys are more frequent for higher-mass halos
at all redshifts, consistent with the $\Lambda$CDM  framework. From the figure, we
can directly see that the number of flybys becomes comparable or larger than the number of mergers for halos $> 10^{11}\,{h^{-1} M_\odot}$
for $z \la 2$.  Such halos are expected to host galaxies -- and the effect of flybys should leave an imprint on
the observable properties.  We also find tentative evidence that flybys are comparable in number to mergers at the
highest redshifts, $z\ga 14$ -- however, the numbers are affected by Poisson error and we can not
conclusively say that flybys dominate mergers at those redshifts.

We checked to see if there are multiple flybys between the same halo pairs
and found that $\sim$ 70\% of the flybys do not recur. About 20\% of flybys eventually become a merger and  $\sim$ 6\%
are repeated flybys. Thus, flybys primarily represent one-off events between two halos. We find that a typical flyby
has a relative velocity $\sim 1.5-2$ times the circular velocity of the primary and 
penetrates to $\sim 30\%$ (up to 10\%) of the primary virial radius. Such close encounters can create a strong perturbtation that
potentially transform the galaxy.

\section{Conclusions}
In this paper we report on a new class of interactions -- flybys, that occur frequently. Most of these flybys are one-off events -- 
one halo delves within the virial radius of another main halo and separates at a later time. 
We find that the number of close flyby interactions is 
comparable to, or even surpasses, the number of mergers for halo masses $\ga 10^{11}\,{h^{-1} M_\odot}$ at $z \la 2$. 
We find that most flybys are one-off events and about 70\% of the flybys do not ever return.  
In general, slow flybys cause a larger perturbation compared to a fast one, and such features can persist even when 
the perturbing halo has moved far away~(Vesperini \& Weinberg, 2000). Flybys are then, a largely-ignored type of interaction that can 
potentially transform galaxies. Unfortunately, most semi-analytic methods of galaxy formation are designed only to use mergers and thus can not account for the
effects of flybys directly.

Although the simulations outlined here pertain strictly to dark matter halo flybys, there are naturally links between 
these results and galaxy flybys. To better simulate the rate of galaxy flybys, we need to populate these halos with 
galaxies. This will allow us to constrain the actual galaxy flyby rate and to statistically compare to  
numerical (e.g., Perez et al. 2006) and observational studies of galaxy-pairs
(e.g., Nikolic et al. 2004, Kewley et al. 2006, Driver et al. 2006). 

\acknowledgments MS would like to thank the organizers for a wonderful conference. This work was conducted in 
part using the resources of the Advanced Computing Center for Research and Education at Vanderbilt 
University. 
We also acknowledge support from the NSF Career award AST-0847696.

\end{document}